\documentclass[preprintnumbers,amsmath,amssymbm,prd]{revtex4}
\usepackage{epsfig}
\usepackage{graphicx}
\usepackage{amssymb}

\begin{document}
\title{Marginally bound resonances of charged massive scalar fields in the background of a charged reflecting shell}
\author{Shahar Hod}
\affiliation{The Ruppin Academic Center, Emeq Hefer 40250, Israel}
\affiliation{ }
\affiliation{The Hadassah Institute, Jerusalem 91010, Israel}
\date{\today}

\begin{abstract}
\ \ \ We study {\it analytically} the characteristic resonance
spectrum of charged massive scalar fields linearly coupled to a
spherically symmetric charged reflecting shell. In particular, we
use analytical techniques in order to solve the Klein-Gordon wave
equation for the composed charged-shell-charged-massive-scalar-field
system. Interestingly, it is proved that the resonant oscillation
frequencies of this composed physical system are determined by the
characteristic zeroes of the confluent hypergeometric function.
Following this observation, we derive a remarkably compact
analytical formula for the resonant oscillation frequencies which
characterize the marginally-bound charged massive scalar field
configurations. The analytically derived resonance spectrum is
confirmed by numerical computations.
\end{abstract}
\bigskip
\maketitle

\section{Introduction}

Recent analytical explorations \cite{Hodup1} of the coupled
Einstein-scalar field equations have revealed the intriguing fact
that spherically symmetric compact reflecting stars \cite{Noteref}
cannot support regular matter configurations made of static
self-gravitating {\it neutral} scalar fields in their exterior
regions. The theorem proved in \cite{Hodup1} has therefore revealed
the interesting fact that these horizonless physical objects share
the no-scalar-hair property with asymptotically flat black holes
\cite{NSH,Notensk,Hodrc,HerR}.

On the other hand, later studies \cite{Hodup2} of the static sector
of the Klein-Gordon wave equation for a scalar field of proper mass
$\mu$ and charge coupling constant $q$ have explicitly demonstrated
that {\it charged} compact reflecting objects have a much richer
phenomenological structure. In particular, it has been proved in
\cite{Hodup2} that, for a compact reflecting shell (or equivalently,
a compact reflecting ball) of electric charge $Q$, there exists a
{\it discrete} set of shell radii,
$\{R_n(\mu,qQ,l)\}_{n=0}^{n=\infty}$ \cite{Notell}, which can
support static regular bound-state configurations made of {\it
charged} massive scalar fields.

The analytical study presented in \cite{Hodup2} has focused on the
physical properties of the {\it static} sector of the composed
charged-shell-charged-massive-scalar-field configurations. It is
important to emphasize that the compact reflecting shell studied in
\cite{Hodup2} was assumed to have a negligible self-gravity [see Eq.
(\ref{Eq2}) below]. This weak gravity assumption implies, in
particular, that the spacetime outside the reflecting shell is well
approximated by the flat space Minkowski metric \cite{Notemsk}.

The main goal of the present paper is to explore the physical
properties of the composed
charged-shell-charged-massive-scalar-field configurations
\cite{Hodup2}. In particular, we shall analyze the resonance
oscillation spectrum $\{\omega_n(\mu,qQ,l,R)\}_{n=0}^{n=\infty}$
which characterizes the stationary \cite{Notestat} bound-state
configurations of the linearized charged massive scalar fields in
the background of the charged reflecting shell. These bound-state
(spatially localized) field configurations are characterized by
proper resonant frequencies which are bounded from above by the
relation
\begin{equation}\label{Eq1}
\omega^2_{\text{field}}<\mu^2\  .
\end{equation}
The inequality (\ref{Eq1}) guarantees that the stationary charged
massive scalar field configurations are characterized by
normalizable radial eigenfunctions which are spatially bounded to
the central charged reflecting shell [see Eq. (\ref{Eq8}) below].

In order to explore the resonance spectrum of the composed
charged-shell-charged-massive-scalar-field system, we shall study in
this paper the physical and mathematical properties of the
Klein-Gordon wave equation for stationary charged massive scalar
fields linearly coupled to a charged compact reflecting shell.
Interestingly, as we shall explicitly show below, the resonant
oscillation frequencies which characterize the marginally-bound
charged massive scalar field configurations can be determined {\it
analytically}.

\section{Description of the system}

We shall explore the dynamics of a charged massive scalar field
$\Psi$ which is linearly coupled to a spherically symmetric charged
reflecting shell. The central supporting shell is assumed to have a
negligible self-gravity \cite{Noteunit,Notewga}:
\begin{equation}\label{Eq2}
M,Q\ll R\  ,
\end{equation}
where $\{M,Q,R\}$ are respectively the mass, electric charge, and
proper radius of the central reflecting shell.

The dynamics of a scalar field $\Psi$ of proper mass $\mu$ and
charge coupling constant $q$ \cite{Noteqm} in the background of a
charged shell is determined by the Klein-Gordon wave equation
\cite{HodPirpam,Stro,HodCQG2,Hodch1,Hodch2}
\begin{equation}\label{Eq3}
[(\nabla^\nu-iqA^\nu)(\nabla_{\nu}-iqA_{\nu})-\mu^2]\Psi=0\  ,
\end{equation}
where $A_{\nu}=-\delta_{\nu}^{0}{Q/r}$ is the electromagnetic
potential of the central charged shell. Substituting into the wave
equation (\ref{Eq3}) the decomposed expression \cite{Noteom}
\begin{equation}\label{Eq4}
\Psi(t,r,\theta,\phi)=\int\sum_{lm}e^{im\phi}S_{lm}(\theta)R_{lm}(r;\omega)e^{-i\omega
t} d\omega\
\end{equation}
for the eigenfunction of the stationary charged massive scalar
field, one finds that the radial scalar eigenfunction $R_{lm}(r)$ is
determined by the ordinary differential equation
\cite{HodPirpam,Stro,HodCQG2,Hodch1,Hodch2}
\begin{equation}\label{Eq5}
{{d} \over{dr}}\Big(r^2{{dR_{lm}}\over{dr}}\Big)+\big[(\omega
r-qQ)^2-(\mu r)^2-K_l\big]R_{lm}=0\ ,
\end{equation}
where $K_l=l(l+1)$ is the angular eigenvalue which characterizes the
angular part $S_{lm}(\theta)$ of the charged scalar eigenfunction
(\ref{Eq4}) \cite{Heun,Abram}. Note that the radial equation
(\ref{Eq5}) is invariant under the symmetry transformation
\begin{equation}\label{Eq6}
qQ\to -qQ\ \ \ \ \text{with}\ \ \ \ \omega\to -\omega\  .
\end{equation}
We shall henceforth assume without loss of generality that
\cite{Notesym}
\begin{equation}\label{Eq7}
qQ>0\  .
\end{equation}

The ordinary radial differential equation (\ref{Eq5}), which
determines the characteristic eigenfunctions $\{R_{lm}(r)\}$ of the
stationary bound-state charged massive scalar field configurations,
should be supplemented by the physically motivated asymptotic
boundary condition
\begin{equation}\label{Eq8}
\Psi(r\to\infty)\sim r^{-1+\kappa}e^{-\sqrt{\mu^2-\omega^2}r}\
\end{equation}
at spatial infinity, where $\kappa\equiv
-{{qQ\omega}/{\sqrt{\mu^2-\omega^2}}}$ [see Eq. (\ref{Eq16}) below].
In the bounded frequency regime $-\mu<\omega<\mu$ [see (\ref{Eq1})],
the large-$r$ asymptotic behavior (\ref{Eq8}) corresponds to
spatially localized (normalizable) configurations of the charged
massive scalar fields which are bounded to the central charged shell
\cite{Notefm}. In addition, the presence of the central reflecting
shell dictates the inner radial boundary condition
\begin{equation}\label{Eq9}
\Psi(r=R)=0\
\end{equation}
on the characteristic scalar eigenfunctions at the surface $r=R$ of
the charged reflecting shell.

The radial differential equation (\ref{Eq5}), supplemented by the
boundary conditions (\ref{Eq8}) and (\ref{Eq9}), determines the {\it
discrete} resonance spectrum
$\{\omega_n(\mu,qQ,l,R)\}_{n=0}^{n=\infty}$ which characterizes the
composed charged-spherical-shell-charged-massive-scalar-field
configurations. Interestingly, as we shall explicitly show in the
next section, this ordinary differential equation is amenable to an
{\it analytical} treatment.

\section{The resonance condition of the stationary charged massive scalar fields in the background of the charged spherical shell}

In the present section we shall use analytical techniques in order
to derive a remarkably compact resonance condition [see Eq.
(\ref{Eq20}) below] for the characteristic resonant oscillation
frequencies $\{\omega_n(\mu,qQ,l,R)\}_{n=0}^{n=\infty}$ of the
composed charged-spherical-shell-charged-massive-scalar-field
system.

It proves useful to define the new radial eigenfunction
\begin{equation}\label{Eq10}
\psi_{lm}=rR_{lm}\  ,
\end{equation}
in terms of which the radial equation (\ref{Eq5}) takes the
characteristic form \cite{Noteomt}
\begin{equation}\label{Eq11}
{{d^2\psi}\over{dr^2}}+\Big[\Big(\omega-{{qQ}\over{r}}\Big)^2-\mu^2-{{l(l+1)}\over{r^2}}\Big]\psi=0\
\end{equation}
of a Schr\"odinger-like ordinary differential equation.
Interestingly, this equation can be expressed in the familiar form
of the Whittaker differential equation (see Eq. 13.1.31 of
\cite{Abram})
\begin{equation}\label{Eq12}
{{d^2\psi}\over{dz^2}}+\Big[-{1\over4}-{{qQ\omega}\over{\sqrt{\mu^2-\omega^2}z}}+{{(qQ)^2-l(l+1)}
\over{z^2}}\Big]\psi=0\ ,
\end{equation}
where we have used here the dimensionless radial coordinate
\begin{equation}\label{Eq13}
z=2\sqrt{\mu^2-\omega^2} r\  .
\end{equation}

The general solution of the Whittaker differential equation
(\ref{Eq12}) can be expressed in terms of the confluent
hypergeometric functions (see Eqs. 13.1.32 and 13.1.33 of
\cite{Abram}) \cite{Notechg}:
\begin{equation}\label{Eq14}
\psi(z)=e^{-{1\over2}z}z^{{1\over2}+\beta}\big[A\cdot
U({1\over2}+\beta-\kappa,1+2\beta,z)+B\cdot
M({1\over2}+\beta-\kappa,1+2\beta,z)\big]\ ,
\end{equation}
where $\{A,B\}$ are normalization constants,
\begin{equation}\label{Eq15}
\beta\equiv \sqrt{(l+{1\over2})^2-(qQ)^2}\  ,
\end{equation}
and
\begin{equation}\label{Eq16}
\kappa\equiv -{{qQ\omega}\over{\sqrt{\mu^2-\omega^2}}}\  .
\end{equation}

The asymptotic radial behavior of the scalar function (\ref{Eq14})
is given by (see Eqs. 13.1.4 and 13.1.8 of \cite{Abram})
\begin{equation}\label{Eq17}
\psi(z\to\infty)=A\cdot z^{\kappa}e^{-{1\over2}z}+B\cdot
{{\Gamma(1+2\beta)}\over{\Gamma({1\over2}+\beta-\kappa)}}z^{-\kappa}e^{{1\over2}z}\
.
\end{equation}
Taking cognizance of the physically motivated boundary condition
(\ref{Eq8}) for the stationary normalizable bound-state resonances
of the composed charged-spherical-shell-charged-massive-scalar-field
configurations, one concludes that the coefficient of the
exponentially exploding term in the asymptotic expression
(\ref{Eq17}) should vanish:
\begin{equation}\label{Eq18}
B=0\  .
\end{equation}
Thus, we conclude that the radial eigenfunction which characterizes
the stationary bound-state resonances of the charged massive scalar
fields in the background of the central charged shell is given by
the compact expression
\begin{equation}\label{Eq19}
\psi(z)=A\cdot e^{-{1\over2}z}z^{{1\over2}+\beta}
U({1\over2}+\beta-\kappa,1+2\beta,2\sqrt{\mu^2-\omega^2} r)\  ,
\end{equation}
where $U(a,b,z)$ is the confluent hypergeometric function of the
second kind \cite{Abram}.

Taking cognizance of the inner boundary condition (\ref{Eq9}), which
is dictated by the presence of the central reflecting shell,
together with the expression (\ref{Eq19}) for the radial scalar
eigenfunction, one obtains the
resonance condition
\begin{equation}\label{Eq20}
U\Big({1\over2}+\sqrt{\ell^2-\alpha^2}+{{\alpha\varpi}\over{\sqrt{1-\varpi^{2}}}},
1+2\sqrt{\ell^2-\alpha^2},2\gamma\sqrt{1-\varpi^2}\Big)=0\
\end{equation}
for the stationary composed
charged-spherical-shell-charged-massive-scalar-field configurations,
where we have used here the dimensionless physical parameters
\begin{equation}\label{Eq21}
\alpha\equiv qQ\ \ \ \ ; \ \ \ \ \ell\equiv l+{1\over2}\ \ \ \ ; \ \
\ \ \gamma\equiv\mu R \ \ \ \ ; \ \ \ \ \varpi\equiv
{{\omega}\over{\mu}}\  .
\end{equation}
The analytically derived resonance equation (\ref{Eq20}) determines
the {\it discrete} family of resonant oscillation frequencies
$\{\varpi_n(\alpha,\gamma,\ell)\}_{n=0}^{n=\infty}$ which
characterize the stationary bound-state configurations of the
linearized charged massive scalar fields in the background of the
central charged reflecting shell.

\section{Upper bound on the characteristic resonant frequencies of the charged massive scalar fields}

In the present section we shall derive a remarkably compact upper
bound on the resonant oscillation frequencies which characterize the
composed charged-shell-charged-massive-scalar-field system.
Substituting
\begin{equation}\label{Eq22}
R=r^{\delta}\Phi\
\end{equation}
into the radial differential equation (\ref{Eq5}), one obtains
\begin{equation}\label{Eq23}
r^2{{d^2\Phi}\over{dr^2}}+2(\delta+1)r{{d\Phi}\over{dr}}+\big[(\omega
r-qQ)^2-(\mu r)^2-l(l+1)+\delta(\delta+1)\big]\Phi=0\  .
\end{equation}

The boundary conditions (\ref{Eq8}) and (\ref{Eq9}), together with
Eq. (\ref{Eq22}), imply that the radial eigenfunction $\Phi(r)$
which characterizes the stationary bound-state scalar configurations
must have (at least) one extremum point in the interval
\begin{equation}\label{Eq24}
r_{\text{ext}}\in (R,\infty)\  .
\end{equation}
In particular, the eigenfunction $\Phi(r)$ is characterized by the
relations
\begin{equation}\label{Eq25}
\{{{d\Phi}\over{dr}}=0\ \ \ \text{and}\ \ \
\Phi\cdot{{d^2\Phi}\over{dr^2}}<0\}\ \ \ \ \text{for}\ \ \ \
r=r_{\text{ext}}\
\end{equation}
at this extremum point. Taking cognizance of Eqs. (\ref{Eq23}) and
(\ref{Eq25}), one deduces that the composed stationary
charged-shell-charged-massive-scalar-field configurations are
characterized by the relation $(\omega r_{\text{ext}}-qQ)^2-(\mu
r_{\text{ext}})^2-l(l+1)+\delta(\delta+1)>0$, or equivalently
\begin{equation}\label{Eq26}
\Big(\omega-{{qQ}\over{r_{\text{ext}}}}\Big)^2>\mu^2+{{l(l+1)-\delta(\delta+1)}\over{r^2_{\text{ext}}}}\
.
\end{equation}
The strongest lower bound on the expression
$|\omega-qQ/r_{\text{ext}}|$ can be obtained by maximizing the r.h.s
of (\ref{Eq26}). In particular, the term
$-\delta(\delta+1)/r^2_{\text{ext}}$ is maximized for $\delta=-1/2$,
in which case one finds from (\ref{Eq26}) the characteristic
inequality
$|\omega-{{qQ}/{r_{\text{ext}}}}|>\sqrt{\mu^2+{{(l+{1/2})^2}/{r^2_{\text{ext}}}}}$,
which implies \cite{Noterou}
\begin{equation}\label{Eq27}
\omega-{{qQ}\over{r_{\text{ext}}}}<-\sqrt{\mu^2+{{(l+{1\over2})^2}\over{r^2_{\text{ext}}}}}\
.
\end{equation}

Taking cognizance of Eqs. (\ref{Eq1}) and (\ref{Eq27}), one finds
that the resonant oscillation frequencies which characterize the
composed stationary charged-shell-charged-massive-scalar-field
configurations are restricted to the regime \cite{Noteubd,Notestg}
\begin{equation}\label{Eq28}
0<\omega+\mu<{{qQ}\over{R}}\  .
\end{equation}
The characteristic inequalities (\ref{Eq28}) can be expressed in the
dimensionless compact form [see Eq. (\ref{Eq21})]
\cite{Notesch,Schw1,Schw2}
\begin{equation}\label{Eq29}
-1<\varpi<-1+{{\alpha}\over{\gamma}}\  .
\end{equation}

\section{The characteristic resonance spectrum of the stationary composed
charged-shell-charged-massive-scalar-field configurations}

The analytically derived equation (\ref{Eq20}) for the
characteristic resonant oscillation frequencies of the stationary
composed charged-shell-charged-massive-scalar-field configurations
can easily be solved numerically. Interestingly, one finds that, for
given values of the dimensionless physical parameters
$\{\alpha,\gamma,\ell\}$ [see Eq. (\ref{Eq21})], the composed
physical system is characterized by a discrete spectrum
\begin{equation}\label{Eq30}
\varpi_{\infty}<\cdots <\varpi_2<\varpi_1<\varpi_0\equiv
\varpi^{\text{max}}<-1+{{\alpha}\over{\gamma}}\
\end{equation}
of resonant oscillation frequencies, where $\varpi_{\infty}\to -1$
[see Eq. (\ref{Eq37}) below].

In Table \ref{Table1} we present the value of the largest resonant
oscillation frequency $\varpi^{\text{max}}(\alpha)$ which
characterizes the composed
charged-shell-charged-massive-scalar-field system for various values
of the dimensionless charge coupling constant $\alpha$. The data
presented in Table \ref{Table1} reveals the fact that the
characteristic resonant oscillation frequency
$\varpi^{\text{max}}(\alpha)$ is a monotonically increasing function
of the physical parameter $\alpha$ \cite{Noteq1}. It is worth
emphasizing the fact that the numerically computed resonant
oscillation frequencies $\varpi^{\text{max}}$ of the composed
charged-shell-charged-field system conform to the analytically
derived compact upper bound (\ref{Eq29}).

\begin{table}[htbp]
\centering
\begin{tabular}{|c|c|c|c|c|c|c|c|}
\hline $\text{\ Charge coupling constant\ \ } \alpha\ $ & \ $\ 1\ $\
\ & \ $\ 3\ $\ \ & \
$\ 5\ $\ \ \ & \ $\ 7\ $\ \ & \ $\ 9\ $\ \ & \ $\ 11\ $\ \ \\
\hline \ $\ \ \varpi^{\text{max}}(\alpha;l=0,\gamma=1)\ $\ \ \ &\ \
-0.8090\ \ \ &\ \ -0.0021\ \ \ &\ \ 0.1612\ \ \ &
\ \ 0.3481\ \ \ &\ \ 0.5366\ \ \ &\ \ 0.7235\ \ \ \\
\hline \ $\ \ \varpi^{\text{max}}(\alpha;l=1,\gamma=1)\ $\ \ \ &\ \
-0.8789\ \ \ &\ \ -0.1511\ \ \ &\ \ 0.0814\ \ \ &
\ \ 0.2928\ \ \ &\ \ 0.4939\ \ \ &\ \ 0.6886\ \ \ \\
\hline
\end{tabular}
\caption{Resonant oscillation frequencies of the composed
charged-spherical-shell-charged-massive-scalar-field system with
$l=0,1$ and $\gamma=1$. We display the largest resonant oscillation
frequency $\varpi^{\text{max}}(\alpha)$ of the charged massive
scalar fields for various values of the dimensionless charge
coupling constant $\alpha$ [see Eq. (\ref{Eq21})]. It is found that
the dimensionless resonant oscillation frequency
$\varpi^{\text{max}}(\alpha)$ is a monotonically {\it increasing}
function of the dimensionless physical parameter $\alpha$. In accord
with our analytical derivation, the resonant oscillation frequencies
which characterize the composed physical system are bounded from
above by the compact relation $\varpi^{\text{max}}<\alpha/\gamma-1$
[see Eq. (\ref{Eq29})].} \label{Table1}
\end{table}

In Table \ref{Table2} we display the value of the largest resonant
oscillation frequency $\varpi^{\text{max}}(\gamma)$ which
characterizes the composed
charged-shell-charged-massive-scalar-field system for various values
of the dimensionless mass-radius parameter $\gamma$. The data
presented in Table \ref{Table2} reveals the fact that the
characteristic resonant oscillation frequency
$\varpi^{\text{max}}(\gamma)$ is a monotonically decreasing function
of the physical parameter $\gamma$ \cite{Noteq2}.

\begin{table}[htbp]
\centering
\begin{tabular}{|c|c|c|c|c|c|c|c|}
\hline $\text{\ Dimensionless shell radius\ \ } \gamma\ $ & \ $\ 1\
$\ \ & \ $\ 3\ $\ \ & \
$\ 5\ $\ \ \ & \ $\ 7\ $\ \ & \ $\ 9\ $\ \ & \ $\ 11\ $\ \ \\
\hline \ $\ \ \varpi^{\text{max}}(\gamma;l=0,\alpha=5)\ $\ \ \ &\ \
0.1612\ \ \ &\ \ -0.1649\ \ \ &\ \ -0.4510\ \ \ &
\ \ -0.5851\ \ \ &\ \ -0.6641\ \ \ &\ \ -0.7168\ \ \ \\
\hline \ $\ \ \varpi^{\text{max}}(\gamma;l=1,\alpha=5)\ $\ \ \ &\ \
0.0814\ \ \ &\ \ -0.2005\ \ \ &\ \ -0.4668\ \ \ &
\ \ -0.5941\ \ \ &\ \ -0.6700\ \ \ &\ \ -0.7209\ \ \ \\
\hline
\end{tabular}
\caption{Resonant oscillation frequencies of the composed
charged-spherical-shell-charged-massive-scalar-field system with
$l=0,1$ and $\alpha=5$. We present the largest resonant oscillation
frequency $\varpi^{\text{max}}(\gamma)$ of the charged massive
scalar fields for various values of the dimensionless mass-radius
parameter $\gamma$ [see Eq. (\ref{Eq21})]. It is found that the
dimensionless resonant oscillation frequency
$\varpi^{\text{max}}(\gamma)$ is a monotonically {\it decreasing}
function of the dimensionless physical parameter $\gamma$. Note that
the characteristic resonant frequencies of the composed physical
system conform to the analytically derived upper bound
$\varpi^{\text{max}}<\alpha/\gamma-1$ [see Eq. (\ref{Eq29})].}
\label{Table2}
\end{table}

Interestingly, as we shall now prove explicitly, the resonance
condition (\ref{Eq20}), which determines the characteristic resonant
oscillation frequencies
$\{\varpi_n(\alpha,\gamma,\ell)\}_{n=0}^{n=\infty}$ of the charged
massive scalar fields in the background of the central charged
reflecting shell, can be solved {\it analytically} in the asymptotic
regime
\begin{equation}\label{Eq31}
\varpi\to -1^{+}\
\end{equation}
of {\it marginally-bound} charged massive scalar field
configurations.

Using the asymptotic approximation (see Eq. 13.5.16 of \cite{Abram})
\begin{equation}\label{Eq32}
U(a,b,x)\simeq
\Gamma({1\over2}b-a+{1\over4})\pi^{-{1\over2}}e^{{1\over2}x}x^{{1\over4}-{1\over2}b}
\cdot\cos[\sqrt{2(b-2a)x}+\pi(a-{1\over2}b+{1\over4})]\ \ \ \
\text{for}\ \ \ \ a\to -\infty\
\end{equation}
of the confluent hypergeometric function, one can express the
resonance condition (\ref{Eq20}) in the form
\begin{equation}\label{Eq33}
\cos\Big(\sqrt{-8\alpha\gamma\varpi}+
{{\alpha\varpi\pi}\over{\sqrt{1-\varpi^2}}}+{1\over4}\pi\Big)=0\ .
\end{equation}
Defining the dimensionless physical parameter
\begin{equation}\label{Eq34}
x\equiv \sqrt{1-\varpi^2}\  ,
\end{equation}
one finds from (\ref{Eq33})
\begin{equation}\label{Eq35}
x={{\alpha}\over{n-{1\over4}+c}}\ \ \ ; \ \ \ n\in\mathbb{Z}
\end{equation}
in the $x\ll1$ regime [see Eqs. (\ref{Eq31}) and (\ref{Eq34})],
where
\begin{equation}\label{Eq36}
c\equiv{{\sqrt{8\alpha\gamma}}\over{\pi}}\  ,
\end{equation}
and the integer $n\gg1$ is the resonance parameter of the stationary
bound-state charged massive scalar field modes.

Taking cognizance of Eqs. (\ref{Eq34}) and (\ref{Eq35}), one obtains
the remarkably compact analytical formula
\begin{equation}\label{Eq37}
\varpi=-\Big[1-{1\over2}\Big({{\alpha}\over{n-{1\over4}+c}}\Big)^2\Big]\
\ \ ; \ \ \ n\gg1
\end{equation}
for the characteristic resonant oscillation frequencies of the
composed charged-shell-charged-massive-scalar-field system in the
regime (\ref{Eq31}) of marginally-bound scalar configurations.

It is worth noting that the analytically derived resonance spectrum
(\ref{Eq37}) implies, in accord with the numerical data presented in
Table \ref{Table1}, that the resonant oscillation frequencies which
characterize the composed charged-shell-charged-massive-scalar-field
configurations are a monotonically increasing function of the
dimensionless charge coupling constant $\alpha$. In addition, the
analytically derived formula (\ref{Eq37}) implies, in accord with
the numerical data presented in Table \ref{Table2}, that the
characteristic resonant oscillation frequencies of the charged
massive scalar fields are a monotonically decreasing function of the
dimensionless mass-radius parameter $\gamma$.

\section{Numerical confirmation}

In the present section we shall verify the validity of the
analytically derived formula (\ref{Eq37}) for the {\it discrete}
spectrum of resonant oscillation frequencies which characterize the
stationary composed charged-shell-charged-massive-scalar-field
configurations. In Table \ref{Table3} we present the resonant
frequencies $\varpi^{\text{analytical}}(n;\alpha,l,\gamma)$ of the
charged fields as obtained from the analytically derived formula
(\ref{Eq37}) in the regime (\ref{Eq31}) of marginally-bound scalar
configurations. We also present the corresponding resonant
frequencies $\varpi^{\text{numerical}}(n;\alpha,l,\gamma)$ of the
charged fields as obtained from a direct numerical solution of the
resonance condition (\ref{Eq20}) which characterizes the composed
charged-shell-charged-massive-scalar-field system. We find a
remarkably good agreement \cite{Noteapg} between the approximated
resonant oscillation frequencies of the marginally-bound charged
massive scalar fields [as calculated from the analytically derived
formula (\ref{Eq37})] and the corresponding exact resonant
oscillation frequencies of the charged fields [as computed
numerically from the resonance condition (\ref{Eq20})].

\begin{table}[htbp]
\centering
\begin{tabular}{|c|c|c|c|c|c|c|c|c|}
\hline \text{Formula} & \ $\varpi(n=3)$\ \ & \ $\varpi(n=4)$\ \ & \
$\varpi(n=5)$\ \ & \ $\varpi(n=6)$\ \ & \ $\varpi(n=7)$\ \ & \
$\varpi(n=8)$\ \ & \ $\varpi(n=9)$
\ \ & \ $\varpi(n=10)$\ \ \\
\hline \ {\text{Analytical}}\ [Eq. (\ref{Eq37})]\ \ &\ -0.98404\ \
&\ -0.98851\ \ &\ -0.99134\ \ &\ -0.99323\ \ &\ -0.99457
\ \ &\ -0.99555\ \ &\ -0.99628\ \ &\ -0.99685\\
\ {\text{Numerical}}\ [Eq. (\ref{Eq20})]\ \ &\ -0.98354\ \ &\
-0.98827\ \ &\ -0.99120\ \ &\ -0.99315\ \ &\ -0.99451
\ \ &\ -0.99551\ \ &\ -0.99625\ \ &\ -0.99683\\
\hline
\end{tabular}
\caption{Resonant oscillation frequencies of the composed
charged-spherical-shell-charged-massive-scalar-field system with
$\alpha=1$, $l=0$, and $\gamma=10$. We display the resonant
oscillation frequencies
$\varpi^{\text{analytical}}(n;\alpha,l,\gamma)$ of the charged
massive scalar fields as obtained from the analytically derived
compact formula (\ref{Eq37}). We also display the corresponding
resonant oscillation frequencies
$\varpi^{\text{numerical}}(n;\alpha,l,\gamma)$ of the charged fields
as obtained from a direct numerical solution of the resonance
condition (\ref{Eq20}) which characterizes the composed stationary
charged-shell-charged-massive-scalar-field configurations. One finds
a remarkably good agreement \cite{Noteapg} between the approximated
resonant oscillation frequencies of the marginally-bound charged
massive scalar fields [as calculated from the analytically derived
formula (\ref{Eq37})] and the corresponding exact resonant
oscillation frequencies of the charged fields [as computed
numerically from the resonance equation (\ref{Eq20})].}
\label{Table3}
\end{table}

\section{Summary and discussion}

Recent analytical studies \cite{Hodup1,Hodup2} of the {\it static}
sector of the Klein-Gordon wave equation have explicitly proved that
spherically symmetric charged compact reflecting objects can support
static bound-state matter configurations made of charged massive
scalar fields. In the present paper we have focused on the
non-static sector of the composed charged-shell-charged-field
system. In particular, we have explored the physical properties of
the {\it stationary} bound-state
charged-shell-charged-massive-scalar-field configurations.

Our main goal was to determine the resonance oscillation spectrum
$\{\omega_n(\mu,qQ,l,R)\}_{n=0}^{n=\infty}$ which characterizes the
Klein-Gordon wave equation for charged massive scalar fields
linearly coupled to the central charged reflecting shell. The main
physical results derived in the present paper are as follows:

(1) We have proved that the resonant oscillation frequencies which
characterize the stationary bound-state configurations of the
charged massive scalar fields in the background of the charged
reflecting shell are restricted to the regime [see Eqs. (\ref{Eq1})
and (\ref{Eq28})]
\begin{equation}\label{Eq38}
-\mu<\omega<\min(\mu,{{qQ}\over{R}}-\mu)\  .
\end{equation}

(2) It was shown that, for given values of the dimensionless
physical parameters $\{qQ,\mu R,l\}$, there exists a discrete
spectrum of resonant oscillation frequencies $\{\omega_n(qQ,\mu
R,l)\}_{n=0}^{n=\infty}$ which characterize the composed
charged-shell-charged-massive-scalar-field system. Interestingly, we
have proved that the resonant frequencies of this composed physical
system are determined by the characteristic zeroes of the confluent
hypergeometric function [see Eq. (\ref{Eq20})]. In particular, it
has been explicitly shown that the characteristic resonance spectrum
of the composed charged-shell-charged-field system is determined by
the analytically derived resonance equation [see Eqs. (\ref{Eq20})
and (\ref{Eq21})]
\begin{equation}\label{Eq39}
U\big({1/2}+\sqrt{(l+1/2)^2-(qQ)^2}+{{qQ\omega}/{\sqrt{\mu^2-\omega^{2}}}},
1+2\sqrt{(l+1/2)^2-(qQ)^2},2\sqrt{\mu^2-\omega^2}R\big)=0\  .
\end{equation}

(3) It was found that the characteristic resonant frequency
$\varpi^{\text{max}}$ [see Eq. (\ref{Eq21})] of the composed
charged-shell-charged-massive-scalar-field system is a monotonically
increasing function of the dimensionless charge coupling parameter
$qQ$ and a monotonically decreasing function of the dimensionless
mass-radius parameter $\mu R$.

(4) We have explicitly proved that the resonant oscillation
frequencies which characterize the stationary bound-state charged
massive scalar fields can be determined {\it analytically} from the
resonance condition (\ref{Eq20}) in the asymptotic regime
$\omega/\mu\to -1^{+}$ [see Eq. (\ref{Eq31})] of marginally-bound
scalar configurations. In particular, we have used analytical
techniques in order to derive the remarkably compact analytical
formula [see Eqs. (\ref{Eq21}), (\ref{Eq36}), and (\ref{Eq37})]
\begin{equation}\label{Eq40}
{{\omega}\over{\mu}}=-\Big[1-{1\over2}\Big({{qQ}\over{n-{1\over4}+\sqrt{8qQ\mu
R}/\pi}}\Big)^2\Big]\ \ \ ; \ \ \ n\gg1
\end{equation}
for the resonant oscillation frequencies which characterize the
marginally-bound stationary charged massive scalar field
configurations.

(5) Finally, we have explicitly shown that the analytically derived
formula (\ref{Eq40}) for the characteristic resonance oscillation
spectrum of the marginally-bound
charged-shell-charged-massive-scalar-field configurations agrees
with direct numerical computations of the corresponding resonant
oscillation frequencies.

\bigskip
\noindent
{\bf ACKNOWLEDGMENTS}
\bigskip

This research is supported by the Carmel Science Foundation. I would
like to thank Yael Oren, Arbel M. Ongo, Ayelet B. Lata, and Alona B.
Tea for helpful discussions.



\begin{thebibliography}{99}
\bibitem{Noteref} Here the term `reflecting star' is used to describe a compact reflecting object which is
characterized by a vanishing scalar field on its outer surface [see
Eq. (\ref{Eq9}) below].
\bibitem{Notell} Here the integer $l$ is the spherical harmonic index which characterizes the charged massive
scalar field configurations [see Eq. (\ref{Eq4}) below].
\bibitem{Notemsk} In particular, the spacetime outside the
compact reflecting shell studied in \cite{Hodup2} was assumed to be
characterized by the flat space metric
$ds^2=-dt^2+dr^2+r^2(d\theta^2+\sin^2\theta d\phi^2)$.
\bibitem{Noteunit} We shall use natural units in which $G=c=\hbar=k_{\text{B}}=1$.
\bibitem{Notewga} Note that the weak gravity assumption (\ref{Eq2})
implies that the spacetime metric outside the reflecting shell is
well approximated by the flat space line element
$ds^2=-dt^2+dr^2+r^2(d\theta^2+\sin^2\theta d\phi^2)$.
\bibitem{Noteqm} It is worth noting that the physical parameters $\{\mu,q\}$ which
characterize the bound-state charge massive scalar field stand
respectively for $\{\mu/\hbar,q/\hbar\}$. These characteristic
physical parameters of the stationary field configuration therefore
have the dimensions of $($length$)^{-1}$.
\bibitem{Noteom} The integer indices $\{l,m\}$ are the spherical harmonic
index and the azimuthal harmonic index of the stationary bound-state
scalar eigenfunction, respectively.
\bibitem{Noteomt} For brevity, we shall henceforth omit the angular harmonic indices $\{l,m\}$
of the stationary scalar field mode.
\bibitem{Notechg} Here $M(a,b,z)$ and $U(a,b,z)$ are respectively the confluent hypergeometric
function of the first kind and the confluent hypergeometric function
of the second kind \cite{Abram}.
\bibitem{Noterou} Taking cognizance of (\ref{Eq1}) and (\ref{Eq7}), one realizes that the alternative inequality
$\omega-{{qQ}/{r_{\text{ext}}}}>\sqrt{\mu^2+{{(l+{1\over2})^2}/{r^2_{\text{ext}}}}}$
cannot be satisfied.
\bibitem{Noteubd} Here we have used the inequality
${{qQ}/{r_{\text{ext}}}}<{{qQ}/{R}}$ for $r_{\text{ext}}>R$ [see Eq.
(\ref{Eq24})]. In addition, we have used here the inequality
$-\sqrt{\mu^2+{{(l+{1\over2})^2}/{r^2_{\text{ext}}}}}<-\mu$.
\bibitem{Notestg} Note that, in the regime $\mu>qQ/2R$, the analytically derived
upper bound $\omega<-\mu+qQ/R$ [see Eq. (\ref{Eq28})] on the
resonant oscillation frequencies of the charged massive scalar
fields is stronger than the upper bound $\omega<\mu$ [see Eq.
(\ref{Eq1})] which characterizes the stationary bound-state massive
scalar field configurations.
\bibitem{Notesch} Note that the Schwinger quantum phenomenon \cite{Schw1} associated with the
pair-production mechanism of oppositely charged particles in the
electric field of the central charged shell is exponentially
suppressed in the regime $E_{\text{shell}}\ll E_{\text{c}}$
\cite{Schw1,Schw2}, where $E_{\text{shell}}=Q/R^2$ and
$E_{\text{c}}\equiv \mu^2/q\hbar$ are respectively the electric
field at the surface of the central charged shell and the critical
electric field for the pair-production mechanism of oppositely
charged particles \cite{Schw1,Schw2}. Thus, this quantum effect is
exponentially suppressed in the $\alpha\ll \gamma^2$ regime [this
strong inequality corresponds to $qQ\ll(\mu R)^2$, see Eq.
(\ref{Eq21})].
\bibitem{Noteq1} It is worth emphasizing that the monotonic dependence of the resonant frequency $\varpi^{\text{max}}$ on
$\alpha$ is an important feature of both the spherically symmetric
$l=0$ field configurations and the non-spherically symmetric $l=1$
field configurations.
\bibitem{Noteq2} It is worth noting that the monotonic dependence of the resonant frequency $\varpi^{\text{max}}$ on
$\gamma$ is an important characteristic of both the spherically
symmetric $l=0$ field configurations and the non-spherically
symmetric $l=1$ field configurations.
\bibitem{Noteapg} It is worth noting that the data presented in Table
\ref{Table3} reveals the interesting fact that the agreement between
the exact resonant oscillation frequencies [as obtained from a
direct numerical solution of the resonance condition (\ref{Eq20})]
and the compact analytical formula (\ref{Eq37}) is quite good even
in the regime $n=O(1)$. This observation is quite surprising since
the analytically derived formula (\ref{Eq37}) for the characteristic
resonant oscillation frequencies of the composed
charged-shell-charged-massive-scalar-field system is formally valid
in the $n\gg1$ regime [or equivalently, in the regime $x\ll1$, see
Eqs. (\ref{Eq31}), (\ref{Eq34}), and (\ref{Eq35})].
\end{thebibliography}
\end{document}